\documentclass[reqno,a4paper,oneside]{amsart}
\pdfoutput=1 
\usepackage[foot]{amsaddr} 
\usepackage{amssymb,latexsym} 
\usepackage{graphicx}
\usepackage{bm}
\usepackage{amsmath}

\usepackage[numbers,sort]{natbib} 
\usepackage{hyperref} 
\usepackage{doi} 

\begin{document}
\title[The discrete REM and one step RSB]{The discrete random energy model and one step replica symmetry breaking}

\author{Bernard Derrida$^{1,2}$}
\address{$^1$Coll\`ege de France\\ 11 place Marcelin Berthelot\\ 75005 Paris\\ France}
\address{$^2$Laboratoire de Physique de l'Ecole Normale Sup\'erieure\\ ENS\\ Universit\'e PSL\\ CNRS, Sorbonne Universit\'e\\ Universit\'e de Paris\\ F-75005 Paris\\ France}
\email{bernard.derrida@college-de-france.fr}

\author{Peter Mottishaw$^3$}
\email{peter.mottishaw@ed.ac.uk}
\address{$^3$SUPA\\ School of Physics and Astronomy\\ University of Edinburgh\\
	Peter Guthrie Tait Road\\ Edinburgh EH9 3FD\\ United Kingdom}

\begin{abstract}
	We solve the random energy model when the energies of the configurations take only integer values. In the thermodynamic limit, the average overlaps remain size dependent and oscillate as the system size increases. While the extensive part of the free energy can still be obtained by a standard replica calculation with one step replica symmetry breaking, it is no longer possible to recover the overlaps in this way.  A possible way to adapt the replica approach is to allow the sizes of the blocks in the Parisi matrix to fluctuate and to take complex values.

\end{abstract}

\maketitle

\section{Introduction}
Replica symmetry breaking  (RSB) invented by Parisi in 1979  to solve   spin glass models in their mean field version \cite{Parisi_1979_Infinite,Parisi_1980_Sequence,Parisi_1980_order,Mezard_1984_Nature,Mezard_1984_Replica}
 has been used since in a large number of contexts including the theory of neural networks, problems of optimization, directed polymers and glasses
\cite{Mezard_1987_Spin,Mezard_2009_Information}. 

One of the first achievements of Parisi's theory was 
to predict the  exact expression of  the extensive part of the free energy of the mean field model of spin glasses introduced by  Sherrington  and Kirkpatrick (the SK model) \cite{Sherrington_1975_Solvable}. After  20 years of effort, this exact expression  for the free energy was confirmed by  rigorous mathematical studies \cite{Guerra_2003_Broken,Guerra_2002_Thermodynamic,Guerra_1996_overlap,Talagrand_2003_Spin,Talagrand_2010_Mean}. 

The other striking   predictions  associated with RSB concern the overlaps, which represent a way of  quantifying the fluctuations of the free energy landscapes \cite{Parisi_1983_Order,Mezard_1985_Random}.   Several   of the predictions  of Parisi's theory, such as  ultrametricity or the Ghirlanda-Guerra relations were also mathematically confirmed \cite{Ghirlanda_1998_General,Panchenko_2013_Sherrington,Guerra_2014_Interpolation} but some of  these proofs are dependent on the Gaussian character of the interactions.

In the present note, we will report on a simple case, the discrete random energy model  (DREM),  for which the standard one step replica symmetry breaking  does predict  the correct (extensive part) of the free energy but fails to give the correct expression for the overlaps. In contrast to the original random energy model, the energies in the DREM are no longer Gaussian but instead take only integer values. The DREM has already been considered to locate the complex zeroes of the partition function \cite{Moukarzel_1991_Numerical} and to investigate phase transitions of the $n$-th moments of the partition function as $n$ varies \cite{Ogure_2005_Exact}. 

For the original  REM as well as for the DREM, the system  consists of  an exponential number  $2^N$  of configurations  ${\mathcal{C}}$ and   the energies $E({\mathcal{C}})$ of these configurations   are independent and identically distributed (i.i.d.) variables with a distribution which scales with the system size  $N$.  The goal then is to determine the average free energy, $\langle F \rangle = -\frac{1}{\beta} \langle \log Z \rangle$, where
\begin{equation}
\langle \log Z(\beta) \rangle  = \left\langle \log\left(\sum_{{\mathcal{C}}=1}^{2^N} \ e^{-\beta E({\mathcal{C}})} \right) \right\rangle  
\label{lZ0}
\end{equation}
\big(where $\langle .\rangle$ denotes an average over disorder, i.e. over  the $2^N$ energies $E({\mathcal{C}})$\big)
and the average overlaps $Y_k$
 which are defined by
\begin{equation}
\left\langle Y_k (\beta)\right\rangle = \left\langle \frac{\sum_{\mathcal{C}} e^{-\beta k E({\mathcal{C}})}
}{ \Big(\sum_{\mathcal{C}} e^{-\beta  E({\mathcal{C}})}\Big)^k}\right\rangle  = \left\langle \frac{Z(k \beta)}{Z(\beta)^k} \right\rangle
\label{Yk-def}
\end{equation}  \big(In the case of random energy models, $\langle Y_k \rangle$ is simply the probability of finding $k$ copies of the system in the same configuration \cite{Derrida_1985_Sample,Mezard_1985_Random,Derrida_1997_random}.\big)

In the original REM \cite{Derrida_1980_Random, Derrida_1981_Random, Bovier_2003_Rigorous,Bovier_2006_Statistical,Mezard_2009_Information}
\begin{equation}
P(E)= \frac{1 }{ \sqrt{\pi N}} e^{-\frac{E^2 }{ N}}  
\label{eq1}
\end{equation}
For this choice  it is known 
that,  in the low temperature phase (the frozen phase)  $\beta> \beta_c= 2 \sqrt{\log2} $ the average free energy (\ref{lZ0}) 
and the overlaps (\ref{Yk-def}) 
are given by, in the large $N$ limit, 
\begin{multline}
\langle \log Z(\beta) \rangle  = N \beta {\sqrt{\log 2}}
 -\frac{\beta }{ 2 \beta_c} \log\left( 4 N \pi \log 2 \right) \\+ \frac{\beta}{ \beta_c} \log \left( \Gamma\left(1- \frac{\beta_c }{ \beta}\right)\right)+\left(1-\frac{\beta}{ \beta_c} \right) \Gamma'(1)
\label{eq2}
\end{multline}

\begin{equation}\langle Y_k(\beta) \rangle = 
\frac{\Gamma(k-\frac{\beta_c }{ \beta})}{ 
\Gamma(1-\frac{\beta_c }{ \beta})\ (k-1)!} 
\label{eq3}
\end{equation}
up to vanishing small corrections in the large $N$ limit
\cite{Gross_1984_simplest,Derrida_1985_Sample,Derrida_1987_Statistical,Ruelle_1987_Mathematical,Derrida_1997_random}.

Our goal in the present work is to see how these expressions are modified when the energies $E({\mathcal{C}})$ are discrete, for example when
\begin{equation}
P(E=\nu) = \left(\begin{array}{c} N \\ \nu \end{array} \right) (1-p)^\nu \ p^{N-\nu} \  \ \ \ \ , \ \ \  \nu=0,1 \cdots N .
\label{binomial}
\end{equation}
In (\ref{binomial}) each energy can be viewed as the sum of $N$ independent binary  random variables which take  value  $0$ with probability $p$ and  $1$ with probability $1-p$ (while in (\ref{eq1}) each energy is a sum of $N$ independent Gaussians).

Our main result  will be  to show that, although the replica approach still predicts the correct extensive part of the free energy,  the expression
(\ref{eq3})
of the overlaps  is no longer valid.
In section \ref{exact},  we will show, by an exact calculation, that the main difference between  the Gaussian (\ref{eq1}) and the discrete energy (\ref{binomial}) cases is that the $O(1)$ finite size corrections of free energy as well as the overlaps  are no longer constant as in (\ref{eq2},\ref{eq3}) but oscillate as   the system size increases, even in the  large 
$N$ limit. As discussed in section \ref{replica}, this raises the question of how to adapt the replica symmetry breaking  calculation to reproduce these periodic dependences.

\section{ How to calculate the free energy and the overlaps}

In this section  we establish the following  three  formulas, valid  for  both the original REM and the DREM (in their Poisson process versions), which    give the average free energy, the negative moments of the partition function and the average overlaps  in terms of the generating function  $\langle e^{-t Z} \rangle$ of the partition function:

\begin{equation}
\langle \log Z(\beta) \rangle   = \int_0^\infty dt \, \frac{e^{-t} - \langle e^{-t Z } \rangle }{t} 
 = \int_0^\infty dt \, \log t \  \left(e^{-t} + \frac{d  \, 
{ e^{\phi(t) }} }{d t}\right) 
\label{lZ}
\end{equation}

\begin{equation}
\langle  Z(\beta)^{n}  \rangle 
  = \frac{1 }{ \Gamma(-n)} \int_0^\infty dt \,  t^{-n-1} \ e^{\phi(t) } 
  = \frac{1 }{ \Gamma(1-n)} \int_0^\infty dt \,  t^{-n} \ \frac{d \,e^{\phi(t)} }{dt} 
 \ \ \ \ \ \text{for} \ \ \ n <0
\label{Zn3}
\end{equation}

\begin{equation}
\langle Y_k (\beta) \rangle =\frac{1 }{\Gamma(k)}  \int_0^\infty dt \, t^{k-1}  \  e^{\phi(t) } \left[    (-)^k \ \frac{d^k \phi(t}{ dt^k}  \right]
\label{Yk}
\end{equation}
where $\phi(t)$ is defined by
\begin{equation}
e^{\phi(t)} =
\langle e^{-t \, Z } \rangle \ . 
\label{phi-def}
\end{equation}

These expressions will be used in section \ref{exact}.
(For $n>0$ alternative expressions     could 
easily be obtained from (\ref{Zn3}) by successive integrations by parts.
We won't use them below. 
Also for positive  $n$, as  $n$ varies, the moments $\langle Z^n \rangle$   exhibit phase transitions  \cite{Gardner_1989_probability,Ogure_2005_Exact,Ogure_2009_analyticity} 

The first two identities (\ref{lZ},\ref{Zn3}) are obviously valid for any positive random variable $Z$.   
Let us now
  establish the third one (\ref{Yk}) for the DREM, i.e. in the case of where the  energies $E$ can only take integer values as in (\ref{binomial}). The derivation can easily be extended to the case of continuous energies.

It is well known, that in the low temperature phase (and in the large $N$ limit), only configurations with  energies  close   enough to the ground state (i.e. such that $E({\mathcal{C}}) - E_\text{ground state} \ll N$) matter \cite{Ruelle_1987_Mathematical}. This implies that,   
 up to vanishing small corrections \cite{Derrida_2015_Finite}, 
one can replace 
  a REM of $2^N$ energy levels distributed according to a distribution $P(E)$  by  a Poisson REM with energies    generated by a Poisson process of density
$$\rho(E) = 2^N P(E) \ . $$

 In the case of the DREM, where the energies take discrete values $\nu$, several configurations ${\mathcal{C}}$ may have the same energy $E({\mathcal{C}}) $ and the partition function is of the from
\begin{equation}
Z = \sum_\nu m_\nu \ e^{-\beta \nu }  
\label{Z-discrete}
\end{equation}
where $m_\nu$ is the number of configurations ${\mathcal{C}}$ such that $E({\mathcal{C}})=\nu$.
Then since
the density
$\rho(E)$ is of the form
$$\rho(E)= \sum_{\nu=-\infty}^\infty r_\nu\  \delta(E-\nu)   \ \ \ \ \ \text{with} \ \ \ \ \ r_\nu= 2^N P(E=\nu)$$
  the numbers $m_\nu$ of configurations with energy $\nu$ are  randomly   distributed according to  Poisson distributions
\begin{equation}
\text{Pro}(m_\nu=m) = \frac{(r_\nu)^m }{ m!} e^{-{r_\nu}}
\label{Pm}
\end{equation}
Moreover these degeneracies $m_\nu$ are independent. 

Then  
the generating function 
{(see (\ref{phi-def}))}
of the partition function $Z$ is given by 
\begin{equation}
\label{eZ5}
\langle e^{-t Z} \rangle = 
\prod_{\nu=-\infty}^\infty \left(\sum_{m=0}^\infty  \frac{r_\nu^m  }{m!}  \ e^{-r_\nu} \exp[-{m} \,  t \, e^{-\beta \nu} ] \right)   
= e^{\phi(t)} 
\end{equation}
with $\phi(t)$  given by
\begin{equation}
\label{phi-discrete}
\phi(t)=
 \sum_\nu r_\nu \Big( \exp[-t e^{-\beta \nu}] -1\Big)
\end{equation}

To obtain the average overlaps $\langle Y_k \rangle$ {defined in (\ref{Yk-def})}  one 
first starts by computing the following average over disorder (i.e. over the $m_\nu$'s)
\begin{equation}
\langle Z(k \beta) \, e^{-t Z(\beta)} \rangle  = \left\langle 
\sum_\nu  m_\nu \, e^{- \beta \nu}
\exp\left[-\sum_{\nu'} m_{\nu'}  \, t \, e^{-\beta \nu'} \right] \right\rangle  
\label{new4}
\end{equation}
Following the same procedure as  for the derivation of   (\ref{phi-discrete}) from (\ref{eZ5}) 
one gets
\begin{equation}
\langle Z(k \beta) \, e^{-t Z(\beta)} \rangle  = (-)^k \   e^{\phi(t)}  \ \frac{d^k \phi(t) }{ dt^k} 
\label{new5} 
\end{equation}

Then rewriting  (\ref{Yk-def})
as
$$ \langle Y_k(\beta) \rangle  = \frac{1 }{(k-1)!} \int_0^\infty d
t \, t^{k-1}  \langle Z(k \beta) \, e^{-t Z(\beta)} \rangle  $$  leads to (\ref{Yk}).

The  validity  of (\ref{Yk}) can  easily be  checked in  the case of the original REM, i.e. when the distribution of energies is continuous. The only difference is {that (\ref{phi-discrete})  becomes}
\begin{equation}
\label{phi-continuous}
\phi(t)= \log \langle e^{-t Z} \rangle
=
\int dE \ \rho(E)  \Big( \exp[-t e^{-\beta E}] -1\Big)
\end{equation}
i.e. that the sum in (\ref{phi-discrete}) is replaced  by an integral.

\section{Continuous versus discrete energies}
\label{exact}
 In the rest of this paper we will consider that each energy $E({\mathcal{C}})$ is the sum of $N$ i.i.d. random variables {as in (\ref{eq1}) or (\ref{binomial})}. Therefore the generating function of each of these energies is of the form
\begin{equation} 
\left\langle e^{-\beta E({\mathcal{C}})} \right\rangle = e^{N \Psi(\beta)}
\label{ge}
\end{equation} 
For example  for the distributions of energies (\ref{eq1}) and (\ref{binomial}) one has respectively
\begin{align}
\Psi(\beta)= & \frac{\beta^2 }{ 4}  \label{Psi1} \\
\Psi(\beta)= & \log\Big(p + (1-p){ e^{-\beta}} \Big) 
\label{Psi2}
\end{align}
From (\ref{ge}), writing that  $\int dE \, P(E) e^{-\beta E} = e^{N \Psi(\beta)} $
one  can see that for large $N$
\begin{equation}
P(E) \simeq  
 \sqrt{\frac{ G''(\frac{E }{ N})}{ 2 N \pi}}
 \ e^{-NG(\frac{E }{ N})}
\label{PE-general}
\end{equation}
 where the function $G(\epsilon)$ can be expressed in a parametric form in terms of $\Psi(\beta)$ 
\begin{equation}
\epsilon=-\Psi'(\beta) \ \ \ \ \ ; \ \ \ \ \ G(\epsilon) = \beta \Psi'(\beta) -\Psi(\beta)
\label{Legendre-1}
\end{equation}
Since $G$ and $\psi$ are Legendre transforms of each other, one also has
\begin{equation}
 \beta=-G'(\epsilon)
 \ \ \ \ \ ; \ \ \ \ \
 \Psi(\beta)=  \epsilon G'(\epsilon) - G(\epsilon)
 \ \ \ \ \ ; \ \ \ \ \
G''(\epsilon) = \frac{1 }{\Psi''(\beta)}
\label{Legendre-2}
\end{equation}
{Here we are assuming that both $G(\epsilon)$ and $\Psi(\beta)$ are C$^2$, i.e. they have at least two continuous derivatives (which they have in the cases (\ref{eq1}) and (\ref{binomial})).}

Because the $2^N$ energy levels are i.i.d., one has that,
 close to the ground state energy,
 $2^N P(E)$ is neither exponentially large in $N$ nor exponentially small in $N$. Therefore if
 $\epsilon_c$ is the minimal solution of 

\begin{equation}
G(\epsilon_c)=\log 2
\label{epsilon_c-def}
\end{equation}
 the extensive  part of the ground state energy is $N \epsilon_c$

In the REM the low temperature phase is dominated by the configurations which have an energy at distance of order 1 from to the ground state energy. Thus  one can replace the distribution $P(E)$ by
an exponential distribution \cite{Ruelle_1987_Mathematical} which approximates the true $P(E)$ near $E=N \epsilon_c$.
Therefore 
\begin{equation}
\rho(E) = 2^N P(E) \simeq  A\,  
e^{ \beta_c(E- N \epsilon_c)} 
\ \ \ \ \ \text{where} \ \ \ \ \ 
A= \sqrt{\frac{ G''(\epsilon_c)}{ 2 \pi N}} 
= \sqrt{\frac{ 1}{2 \pi N \Psi''(\beta_c)} }
\label{PoissonP}
\end{equation}
and $\beta_c$ is defined by
\begin{equation}
\beta_c= -G'(\epsilon_c)
\label{beta_c-def}
\end{equation}

Note that (\ref{Legendre-1},\ref{Legendre-2},\ref{epsilon_c-def},\ref{beta_c-def})   
\begin{equation}
\beta_c \Psi'(\beta_c) - \Psi(\beta_c) = \log 2 \ . 
\label{saddle1}
\end{equation}
This relation will appear in section \ref{replica} when we will look at the replica approach.

Let us now discuss the cases of continuous energies and discrete energies separately:
\subsection{Continuous energies {as in (\ref{eq1})}}

For continuous energies,  the integral in (\ref{phi-continuous}) can be performed exactly  for $\rho(E)$ given by (\ref{PoissonP}) and one gets
\begin{align}
\phi(t)=- \frac{A  \ e^{-N \beta_c \epsilon_c}} {\beta_c} \ t^\frac{\beta_c }{ \beta} \ \Gamma\left(1-\frac{\beta_c }{\beta} \right)
\label{phi1}
\end{align}
and one gets  for  the average free energy (\ref{lZ})
and for the negative moments of the partition function
(\ref{Zn3})
\begin{multline}
\langle \log Z \rangle 
=-N \beta \epsilon_c + \frac{\beta }{ \beta_c} \log \left(\frac{A }{\beta_c} \right) \\ +\frac{\beta }{\beta_c} 
\log\left( 
\Gamma\left(1-\frac{\beta_c }{ \beta} \right)\right) + \left(1-\frac{\beta }{\beta_c} \right) \Gamma'(1)
+o(1)
\label{lZ-cont}
\end{multline}

\begin{equation}
\label{Zn-cont} 
\langle Z^n 
\rangle \simeq \frac{\Gamma \left(1- n \frac{\beta }{ \beta_c} \right) }{ \Gamma(1-n)} \ \left[ \frac{A }{\beta_c} \ \Gamma\left(1- \frac{\beta_c }{ \beta} \right) \right]^\frac{n \beta }{ \beta_c} 
 \ e^{-N n \beta \epsilon_c} 
\end{equation}

It is easy to check using (\ref{PoissonP})  that for $G(\epsilon)=-\epsilon^2$, one has $\epsilon_c= - \sqrt{\log2}, \beta_c =2 \sqrt{\log2}$ that and (\ref{lZ-cont}) reduces to (\ref{eq2}). (Also, not surprisingly  
one can recover (\ref{lZ-cont}) from (\ref{Zn-cont})
in the limit $n \to 0^-$.)
Therefore (\ref{lZ-cont}) gives the generalization of (\ref{eq2}) to other   continuous distributions of energies $G(\epsilon)$.

For the overlaps, using (\ref{phi1}) into (\ref{Yk}) leads to 
\begin{equation}
\langle Y_k(\beta) \rangle = 
\frac{\Gamma(k-\frac{\beta_c }{ \beta}) }{
\Gamma(1-\frac{\beta_c }{ \beta})\ (k-1)!} 
\label{Yk-cont}
\end{equation}
which is identical to (\ref{eq3}). 

It is remarkable, though expected from   the fact that $\rho(E)$ can be replaced by an exponential (\ref{PoissonP})  that the average overlaps, in the large $N$ limit, do not depend on the details of the distribution $P(E)$.

\subsection{Discrete energies as in (\ref{binomial})} 
 When the energies take only integer values,
the parameters $r_\nu$  in (\ref{Pm}) are given by 
 $$r_\nu= 2^N \, P(E = \nu) =  A\,  {e^{-N \beta_c \epsilon_c}}\  e^{\beta_c \nu}$$
Then  $\phi(t)$ in (\ref{phi-discrete}) can be written as 
\begin{align}
\phi(t)=
- \frac{A  \ 
 {e^{-N \beta_c  \epsilon_c}}}{\beta_c}
 \  t^\frac{\beta_c }{ \beta}  \ w(t)  
\label{phi-discrete-1}
\end{align}
where 
\begin{equation}
w(t)= \beta_c \sum_{\nu=-\infty}^\infty t^{-\frac{\beta_c }{ \beta} } \, e^{\beta_c \nu}\Big(1-  \exp[-t e^{-\beta \nu}]\Big) 
\label{w1}
\end{equation}
Using the Poisson summation formula 
$$ \sum_n f(n) = \sum_q  \int dz f(z) e^{i 2 \pi q z} $$
one can also write $w(t)$ as
\begin{equation}
w(t)= \sum_{q=-\infty}^\infty  w_q \,  t^\frac{2 i \pi q }{ \beta} \  \ \ \ \ \text{where} \ \ \ \ w_q =-\frac{\beta_c }{ \beta}  \ \Gamma\left( - \frac{\beta_c + 2 i \pi q }{ \beta} \right) 
\label{w2}
\end{equation}

\begin{figure}[h]
\centerline{\includegraphics[width=0.9\linewidth]{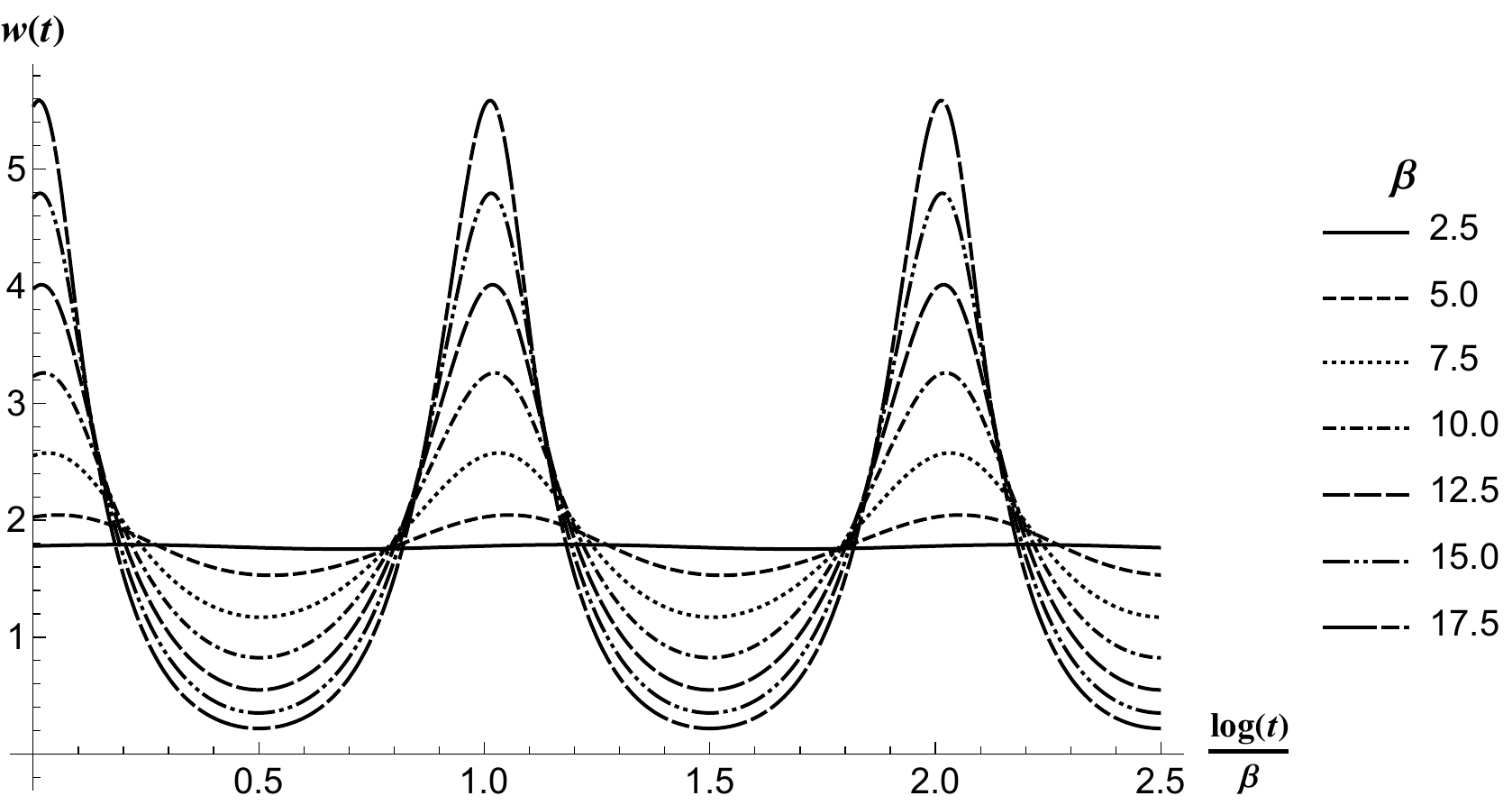}}
\caption{\small $ w(t)$ versus $(\log t)/\beta$ for $\beta_c= \beta/2$ and  $\beta=2.5 , 5 , 7.5 \cdots 17.5$. The amplitude of the oscillations increase with $\beta$.}
\label{Fig1}
\end{figure}

The main difference with the case of continuous energies (\ref{phi1})  is that the function $w(t)$ in (\ref{phi-discrete})
is no longer constant.
It is in fact  easy to check in (\ref{w1}) that
\begin{equation}
w(t) = w(  e^\beta \, t)
\label{w3}
\end{equation}
i.e. is $w(t)$ a periodic function of $\log t/ \beta$ (see Figure \ref{Fig1})

Starting from (\ref{lZ},\ref{Zn3}) and (\ref{Yk})   and making the change of variable 
$$t \to \left(\frac{A }{ \beta_c}\right)^{-\frac{\beta }{ \beta_c}} \  e^{N \beta \epsilon_c} \ \tau$$
one gets for the free energy
\begin{multline} \langle \log Z \rangle  = 
-N \beta \epsilon_c + \frac{\beta }{ \beta_c} \log\left( \frac{A }{ \beta_c}\right) + \Gamma'(1) \\ 
+\int_0^\infty  d \tau \ \log(\tau) \frac{d }{ d \tau} \left[\exp\left(- \tau^\frac{\beta_c }{ \beta} \, w(B \tau) \right) \right] + o(1)
\label{lZ-discrete}
\end{multline}

 for the negative moments of $Z$
\begin{equation}
\langle Z^{n} \rangle  \simeq
\frac{1 }{ \Gamma(1-n)} e^{-n\beta \epsilon_c N } \left(\frac{A }{\beta_c}\right)^{\frac{\beta }{ \beta_c} n } \ \int_0^\infty d \tau \,  \tau^{-n} \ \frac{d }{ d \tau} \left[ \exp[-\tau^\frac{\beta_c }{ \beta} w(B \, \tau )]  \right] 
\label{Zn-discrete}
\end{equation}

and for the overlaps
\begin{equation}
\langle Y_k(\beta) \rangle 
 = \frac{(-)^{k-1} }{ (k-1)!} \int_0^\infty d\tau \, \tau^{k-1} \   \exp\left[-\tau^\frac{\beta_c }{ \beta} w(B \tau) \right] \  \ \frac{d^k \Big( \tau^\frac{\beta_c }{ \beta} \, w(B \tau) \Big) }{ d\tau^k} 
\label{Yk-discrete}
\end{equation}
where 
\begin{equation}
B=\left(\frac{ \beta_c }{ A }\right)^{\frac{\beta }{ \beta_c}} e^{N \beta \epsilon_c} 
\label{B-def}
\end{equation}
with $A$ given in (\ref{PoissonP}).
These expressions (\ref{lZ-discrete},\ref{Zn-discrete},\ref{Yk-discrete},\ref{B-def}) are  to be compared to (\ref{lZ-cont},\ref{Zn-cont},\ref{Yk-cont}).  Because (see (\ref{w3})) the function $w(t)$ is a periodic function of $\frac{\log t }{ \beta}$, the overlaps (\ref{Yk-discrete}) as well as the contributions of order 1 (i.e. the last terms in (\ref{lZ-discrete}) and  the last factor in (\ref{Zn-discrete})) are periodic functions of $(\log B)/\beta$.

\begin{figure}[h]
	\centering
\includegraphics[width=0.9\linewidth]{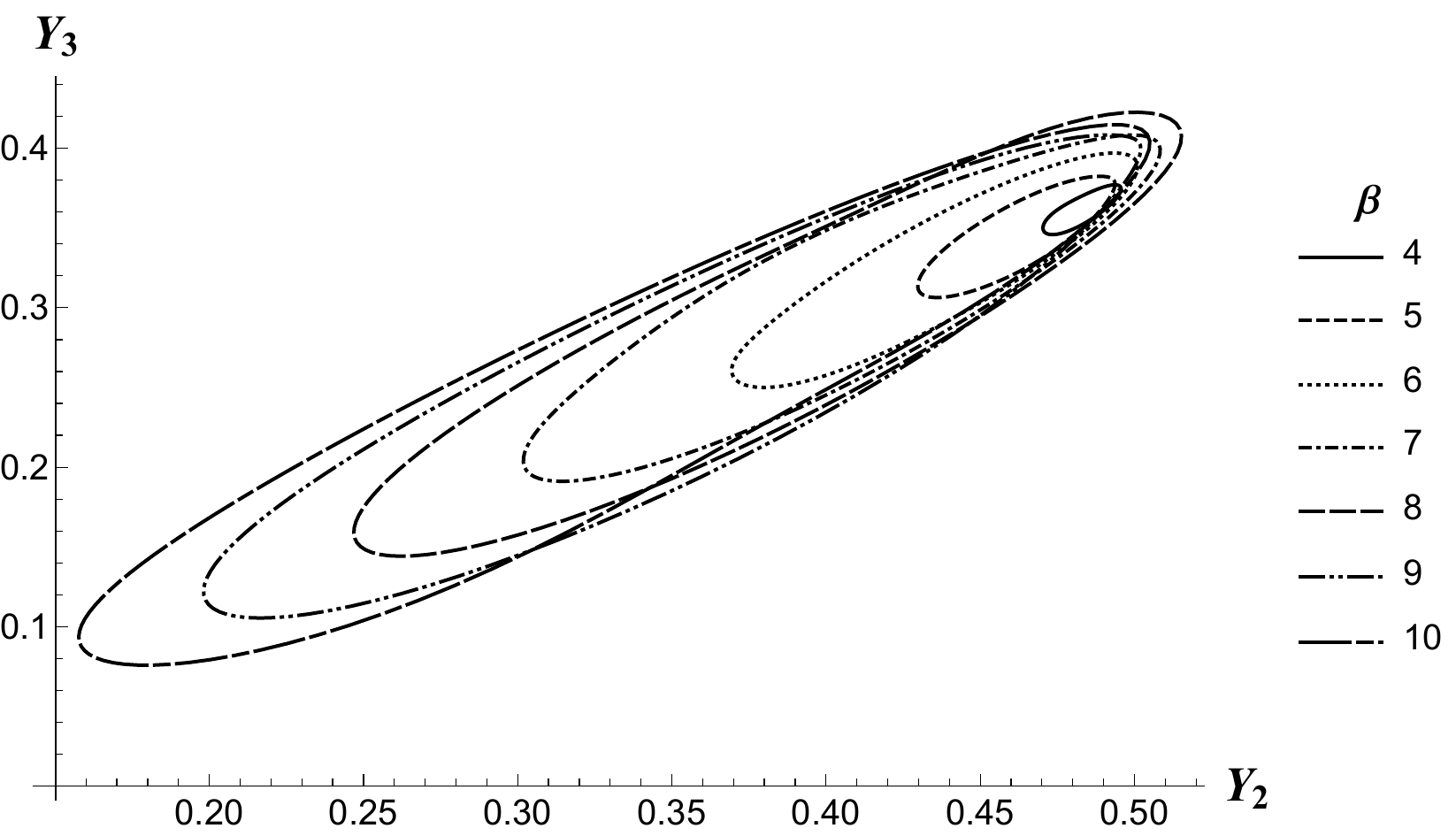}
\caption{
$Y_3$ versus $Y_2$  given by (\ref{Yk-discrete}) when the system size increases at a fixed value of the ratio $\frac{\beta_c }{ \beta}=\frac{1 }{ 2}$. The various ellipses correspond to
 several choices of  $\beta=4,5,\cdots 10$. As the system size increases, $Y_2$ and $Y_3$ vary   when the energies are discrete (\ref{Yk-discrete}). On the contrary for continuous energies, the plots would reduce to a single point: $Y_2=\frac{1 }{ 2}, Y_3=\frac{3 }{ 8}$ (see(\ref{eq3}))  independent of $\beta$ and of the system size (in the limit of large system sizes).}
\label{Fig2}
\end{figure}

The fact that the overlaps remain size dependent  through the parameter $B$  in (\ref{B-def}) as the system size increases can be illustrated  by the fact that in the plane $Y_2,Y_3$, expression (\ref{Yk-discrete})  leads  to closed loops instead of a single point for a fixed choice of $\beta_c$ and $\beta$ (see Figure \ref{Fig2})).

\section{The replica approach}
\label{replica}

In this section we try to see how the exact expressions (\ref{lZ-discrete},\ref{Zn-discrete},\ref{Yk-discrete}) in the discrete energy case and  the corresponding expressions (\ref{lZ-cont},\ref{Zn-cont},\ref{Yk-cont})  in  the continuous energy case   could be recovered by Parisi's replica approach.

\subsection{The  moments of the partition function}
In the replica calculation one usually starts by the calculation of the positive integer moments of the partition function. To do so, 
 one has  (see (\ref{phi-discrete}) or (\ref{phi-continuous})) 
\begin{equation}
\phi(t) =  \sum_{\mu \ge 1}\frac{ (-t)^\mu  }{ \mu!}  \ \langle Z( \beta \mu) \rangle
\label{phiZ}
\end{equation}
and from  (\ref{phi-def})  

$$\langle e^{-t Z} \rangle = \exp\left[ \sum_{\mu \ge 1} \frac{(-t)^\mu   }{ \mu!} \langle Z(\beta \mu ) \rangle \right]$$
so that \cite{Derrida_1981_Random,Dotsenko_2011_Replica}
\begin{align}
\langle Z^n(\beta) \rangle
 &
 =	\sum_{r\ge 1} \frac{n! }{ r!} 
\sum_{\mu_1 \ge 1}  \frac{\langle Z(\beta \mu_1)\rangle }{ \mu_1!}
\cdots \sum_{\mu_r \ge 1}  \frac{\langle Z(\beta \mu_r)\rangle }{ \mu_r!} \ \delta_{\mu_1 + \cdots \mu_r-n} 
\nonumber \\
&  =	\sum_{r\ge 1} \frac{n! }{ r!}  2^{N r} 
\sum_{\mu_1 \ge 1}  
\cdots \sum_{\mu_r \ge 1}  \frac{e^{N[\Psi(\beta \mu_1 ) + \cdots \Psi(\beta \mu_r )]} }{  \mu_1! \times  \cdots  \times \mu_r!} \ \delta_{\mu_1 + \cdots \mu_r-n} 
\label{Zn}
\end{align}
Each term represents a partition of the $n$ replicas into $r$ blocks of sizes $\mu_1, \cdots \mu_r$.

\subsection{The overlaps}
To obtain the overlap  one has from (\ref{phiZ})
$$\frac{d^k \phi(t) }{ dt^k}  =  (-)^k \sum_{n=0}^\infty \frac{(-t)^n \,  }{ n! } \langle Z((n+k)\beta) \rangle$$
and from (\ref{phi-def}) 
$$e^{\phi(t)}= \sum_{n=0}^\infty \frac{(-t)^n }{ n!} \langle Z(\beta)^n \rangle$$
Then  as in (\ref{new5}) 
\begin{equation}
	\langle Z(k \beta) e^{-t Z(\beta)} \rangle = (-)^k \left(\frac{d^k \phi(t) }{ dt^k}\right)  \ e^{\phi(t)}
\label{new1} 
\end{equation}
one gets by expanding in powers of $t$
\begin{align}
\langle  Z(k \beta) \, Z(\beta)^{n-k}  \rangle 
& =  (n-k)! \sum_{\mu=k}^{n} \frac{ \langle Z(\mu \beta) \rangle \, \langle Z(\beta)^{n-\mu}\rangle  }{ (\mu-k)! \, (n-\mu)!}
\nonumber \\
& =  (n-k)! \sum_{\mu=0}^{\infty} \frac{ \langle Z(\mu \beta) \rangle \, \langle Z(\beta)^{n-\mu}\rangle  }{ (\mu-k)! \, (n-\mu)!}
\label{new3}
\end{align} 
where in the last line we simply have extended the sum to the range 0 to infinity, having in mind that   factorials of negative integers are infinite.

This expression is also valid in the special case $k=1$ so that one can write
\begin{equation}
\label{overlap}
 \frac{\langle Z(k \beta) \, Z^{n-k} (\beta) \rangle }{ \langle Z^n(\beta)\rangle} =  {\sum_\mu W(\mu)
  \frac{(n-k)! \, (\mu-1)! }{ (n-1)! \, (\mu-k)!}  }
\end{equation}
where the weights  $W(\mu)$ are defined as
$$W(\mu)=\frac{\langle Z(\mu \beta ) \rangle \, \langle Z(\beta)^{n-\mu}\rangle }{ (\mu-1)! \ (n-\mu)! } \ \left( \sum_{\mu'}
\frac{\langle Z(\mu' \beta) \rangle \, \langle Z(\beta)^{n-\mu'}\rangle \  }{ (\mu'-1)! \ (n-\mu')!} \right)^{-1}$$
These expressions will allow us to obtain the overlaps \ref{Yk-def} using,
$$\langle Y_k(\beta) \rangle= \lim_{n \to 0} \frac{\langle Z(k \beta) \, Z^{n-k} (\beta) \rangle }{ \langle Z^n(\beta)\rangle}$$
but as usual in the replica calculation, the question is to analytically continue  expressions like (\ref{Zn}) or (\ref{overlap})  to non-integer (positive or negative) values of $n$.
To do so we will need to   allow the sizes of  the blocks $\mu$ in (\ref{Zn})  and (\ref{overlap})  to take  non integer values.  

\subsection{The continuous energies case}
\label{cont}
Following Parisi's original approach \cite{Parisi_1980_Sequence} we assume that, for non-integer $n$,  in the limit of a large system size ($N \to \infty$), the multiple sum in (\ref{Zn}) is dominated by a single term, with all the $\mu_i$ equal to a single value $\mu$ and $r=\frac{n }{ \mu}$ so that all blocks have the same size $\mu$.  As he also did, we let this common value $\mu$  be a real number (instead of an integer) and we  also assume that the  contribution to the sum in (\ref{Zn}) is {\it minimum} (in contrast with standard saddle point methods where one would need to find the {\it maximum}) at this  value of $\mu$. In other words, for large $N$ we replace (\ref{Zn}) by

\begin{equation}
 \langle Z^n \rangle = {\mathcal{P}} \times \min_\mu 
 \left[ \frac{n!  }{ (\frac{n }{ \mu })!  \ (\mu !)^\frac{n }{ \mu} } \   
\exp\left[ N \frac{n }{ \mu}\Big( \log 2 +  \Psi(\beta \mu)\Big) \right]  \right] 
\label{Zn-replica}
\end{equation}
where the prefactor ${\mathcal{P}}$ represents the "putative" contributions to  the multiple sums  due to the neighbourhood of the extremum where all the $\mu_i$ are equal to $\mu$. The difficulty in determining the prefactor comes from the fact that once we let  $r$ and the $\mu_i$ take  real instead of  integer values in (\ref{Zn}), it is not clear how to estimate the contribution of the fluctuations in the neighbourhood of the extremum. 

The extremum over $\mu$  is given by the solution $\mu_0$ of 
\begin{equation}
\Psi(\beta \mu) + \log 2- \beta \mu \Psi'(\beta \mu)  = 0 \ .
\label{saddle2}
\end{equation}
Then from (\ref{saddle1}), one can see that 
\begin{equation}
\mu_0=\frac{\beta_c }{ \beta} 
\label{mu}
\end{equation}

On the other hand,  
 using (\ref{mu}),  the expression of $A$ (\ref{PoissonP})  and the fact that $\epsilon_c= - \Psi'(\beta_c)= -(\log 2 + \Psi(\beta_c)) /\beta_c$ (see (\ref{Legendre-1},\ref{epsilon_c-def})) 
the exact expression  (\ref{Zn-cont}) can be rewritten as 
\begin{multline}
\label{Zn-cont-1} 
\langle Z^n 
\rangle \simeq 
\left[\frac{1 }{ 2 \pi  N \Psi''(\beta \mu_0)  \beta^2}\right]^\frac{n  }{ 2 \mu_0}  \
\frac{\Gamma \left(1-  \frac{n }{ \mu_0} \right) }{ \Gamma(1-n)} \
\left[\frac{\Gamma\left(1- \mu_0 \right) }{ \mu_0 }\right]^\frac{n }{ \mu_0}
\\  \times
\exp\left[ N \frac{n }{ \mu_0}\Big( \log 2 +  \Psi(\beta \mu_0)\Big) \right]  
\end{multline}

Clearly the exponential  $N$ dependence is the same   in the exact  (\ref{Zn-cont-1}) and  the replica 
(\ref{Zn-replica}) expressions.  Understanding the prefactor in (\ref{Zn-cont-1})
or the $O(1)$ corrections in  (\ref{lZ-cont})
 using the replica approach 
 is more problematic.
There are  however some    similarities between the exact and the replica expressions.  For example  one could assume that  factorials in (\ref{Zn-replica})  can be replaced by Gamma functions in
 (\ref{Zn-cont-1}): $m! \to \frac{1}{ \Gamma(1-m)}$. Then one would need to argue  that the remaining factors come from Gaussian integrals near the saddle point so that
\begin{equation}
{\mathcal{P}}= 
\left[\frac{1 }{ 2 \pi  N \Psi''(\beta \mu_0)  \mu_0^2 \beta^2}\right]^\frac{n  }{ 2 \mu_0}  \
\label{cal-P}
\end{equation}

In fact there is a way to recover this prefactor by using an integral representation \cite{Campellone_2009_Replica,Derrida_2015_Finite}
, but  this is not strictly speaking a replica calculation because from the very start one tries to compute non-integer moments of the partition function. We won't discuss it here.

To obtain  the average overlaps in the replica  approach one can  rewrite (\ref{overlap}) as
\begin{equation}
\label{overlap1}
\langle Y_k(\beta) \rangle= \frac{\langle Z(k \beta) \, Z^{n-k} (\beta) \rangle }{ \langle Z^n(\beta)\rangle} =  {\sum_\mu W(\mu)
  \frac{\Gamma(1-n) \, \Gamma(k-\mu) }{ \Gamma(k-n) \, \Gamma(1-\mu)}  }
\end{equation}
where we have used the fact that a generalization of the combinatorial factors in (\ref{overlap}) to non-integer values of $n$ and $\mu$  can be written in terms of Gamma functions:

$$\frac{(n-k)!\, (\mu-1)! }{ (n-1)! \, (\mu-k)! } = \prod_{p=1}^{k-1}\left(\frac{p- \mu }{ p-n}\right) = \frac{\Gamma(1-n) \Gamma(k-\mu)}{ \Gamma(k-n) \, \Gamma(1-\mu) } $$
Comparing the exact result  (\ref{Yk-cont},\ref{Yk-def})
$$\langle Y_k(\beta) \rangle = 
\frac{\Gamma(k-\frac{\beta_c }{ \beta}) }{ 
	\Gamma(1-\frac{\beta_c }{ \beta})\ (k-1)!} $$
with the replica expression (\ref{overlap1}), in the $n \to 0$ limit,  we see that they agree provided that we choose
$$W(\mu)= \delta\left(\mu-\frac{ \beta_c }{ \beta}\right)$$
In other words the size $\mu$ of the blocks is fixed and takes the value $\mu_0$ already obtained in the calculation of the non-integer moments of $Z$ (see (\ref{mu})).

\subsection{The discrete energies case}
\label{disc1}

For discrete energies the simplest procedure would be to repeat the replica analysis   we did above in  section \ref{cont}   (i.e. consider that the sum in (\ref{Zn}) is dominated by a single term where all the $\mu_i$'s are equal). This would lead to the expression (\ref{Zn-replica}) allowing   to recover  the right $N$-dependence  in the exponential of (\ref{Zn-discrete}) and therefore the right extensive part of the free energy.  However considering that   the sum over $\mu$ in the replica expression for the overlaps (\ref{overlap1}) reduces to a single term would lead to the same result as the continuous case (\ref{Yk},\ref{Yk-cont}) in contradiction with the exact expression (\ref{Yk-discrete}).

On the other hand, as  $\Psi(\beta)$ in (\ref{Psi2}) is a periodic function ($\Psi(\beta)= \Psi(\beta+ 2 i \pi)$), it is clear that if $\mu_0=\frac{\beta_c }{ \beta}$ is solution of (\ref{saddle2})  then all
\begin{equation}
\mu_q= \frac{\beta_c }{ \beta} + \frac{ 2 i \pi q }{ \beta} \ \ \ \ \ \ \text{with}
   \ \ q \in    \mathbb{Z}
\label{muk}
\end{equation}
are also saddle points at the same height as $\mu_0$. Therefore in (\ref{Zn}), one could keep the same exponential $N$-dependence by allowing the $\mu_i$'s to take  all possible values $\mu_q$.  This gives the correct extensive part of the free energy but as in the continuous energy case, we don't have a clear way of  understanding  the prefactor in (\ref{Zn-cont-1}) using replicas, we did not succeed to understand it in the discrete energies case either.

For the overlap, however, we are  going to see now that one   can reproduce the exact expressions (\ref{Yk-discrete}) if one allows the  values of $\mu$ in (\ref{overlap1}) to fluctuate and to take complex values.
From (\ref{w2}) we have 
$$
\frac{d^k \Big( \tau^\frac{\beta_c }{ \beta} \, w(B \tau) \Big) }{ d\tau^k}
=(-)^{k+1} \sum_{q=-\infty}^\infty  \tau^{\frac{\beta_c + 2 i \pi q }{ \beta}-k} \  B^\frac{2 i \pi q }{ \beta} \ \Gamma\left( k-\frac{\beta_c +2 i \pi q }{ \beta} \right) 
$$
Then  inserting this expression into (\ref{Yk-discrete})  and using (\ref{muk}) one gets the exact expression
\begin{equation}
\label{Yk-complex}
\langle Y_k(\beta) \rangle = \sum_{q=-\infty}^\infty W_q \ \frac{\Gamma(k-\mu_q) }{ \Gamma(k) \, \Gamma(1-\mu_q) }
\end{equation}
where
\begin{equation}
\label{Wq}
W_q= B^\frac{2 i \pi q }{ \beta} \, \frac{\beta_c }{ \beta} \, \Gamma(1-\mu_q)  \int_0^\infty \tau^{\mu_q-1} \, \exp\Big[-\tau^{\frac{\beta_c }{ \beta}} w(B \tau) \Big] d \tau
\end{equation}
This is precisely the same form as the replica expression (\ref{overlap1}) (in the $n \to 0$ limit). Therefore  (\ref{Yk-complex}) shows that one way to interpret the exact expression (\ref{Yk-discrete}) within the replica approach is to allow the sizes of the blocks to fluctuate and to take complex values (with also complex weights $W_q$).

\section{Conclusion}
In the present paper we have obtained exact expressions for the overlaps and for the finite size corrections for the random energy model with discrete energies.
Although the standard replica  broken symmetry calculation does give the correct extensive part of the free energy, it fails to predict the overlaps.
In fact, in the thermodynamic limit,  the exact expression (\ref{Yk-discrete}) of the overlaps   oscillates as   the system size $N$ increases. Trying to interpret these oscillations   using replicas,  we saw   that one needs to allow the sizes of the blocks in the Parisi ansatz to fluctuate and to take complex values.

\begin{figure}[h]
	\includegraphics[width=1.0\linewidth]{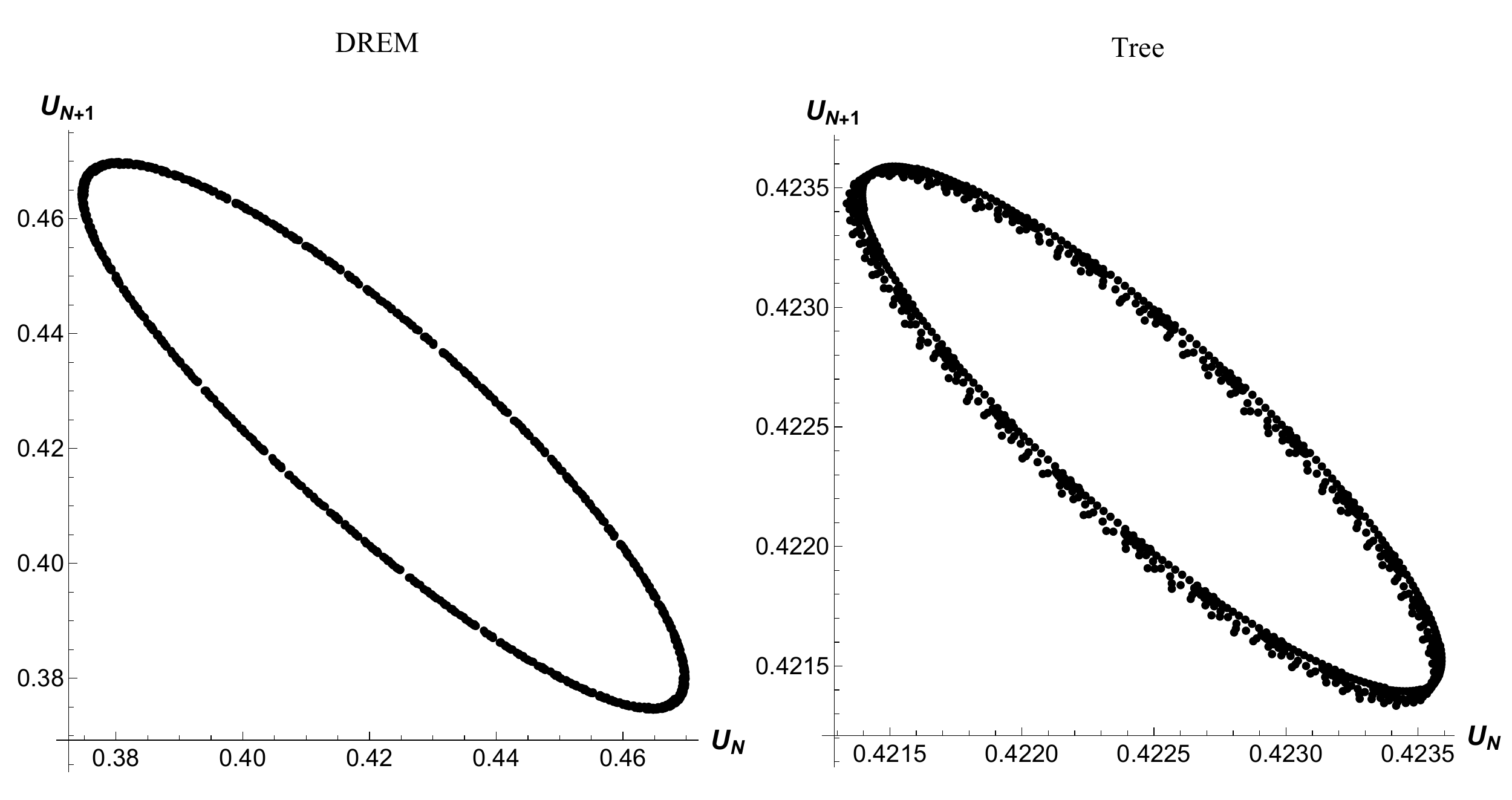}
	\caption{
		We plot $U_{N+1}$ versus $U_N$, where $U_N = E_N - E_{N-1} -\frac{\gamma }{ N \beta_c}$ for the DREM on the left and for the directed polymer on the tree on the right. The parameters chosen are $p=.1, \epsilon_c=.42251, \beta_c=2.5097$ and $ 500< N < 1500$. We see clearly the oscillations. In absence of oscillations, the two graphs would reduce to a single value at $\epsilon_c$.
		One can notice that for the tree the amplitude of the oscillations is much smaller than for the DREM.
	}
	\label{Fig3}
\end{figure}

That  $\langle Y_k \rangle \neq 1 $ at zero temperature is not a surprise because it is expected that the ground state may be degenerate at zero temperature \cite{Sasaki_2002_Temperature,Krzakala_2001_Discrete}. What we have shown here is that  even at finite temperature, the standard one step RSB  does give the correct free energy but does not give the correct overlaps. That the oscillations appear both in overlaps and in the $O(1)$ corrections of the free energy makes a lot of sense  because in the multi-valley picture of the replica symmetry breaking, a variation  of $O(1)$ of  the free energy of a valley  would  change the overlaps. It would be interesting to see whether a similar phenomenon could appear in  more sophisticated  models like the binary perceptron \cite{Gardner_1988_Optimal,Gardner_1989_Three,Krauth_1989_Storage} or the K-sat problem  \cite{Monasson_1997_Statistical,Mezard_2002_Analytic}  for which  the energy spectrum is discrete and the ground state is degenerate.   One case at least for which the oscillations of the $O(1)$ in the free energy is  visible
(see Figure \ref{Fig3})  is
 the case of the directed polymer on a  tree \cite{Derrida_1988_Polymers} where the energies $\epsilon_b$ of the bonds have a binary distribution 
$\epsilon_b=1 $ with probability $1-p$ and $\epsilon_b=0$ with probability $p$.  For the DREM as well for the tree  the ground state energy $E_N$ is of the form 
\begin{equation}
E_N = N \epsilon_c + \frac{\gamma }{  \beta_c}  \log N + e_N + o(1) 
\label{EN}
\end{equation}
with $\gamma =\frac{1}{2}$ for the DREM and $\gamma = \frac{3 }{ 2}$ for the tree \cite{Bramson_1983_Convergence,Bramson_1978_Maximal,Majumdar_2000_Extremal}.
According to (\ref{lZ-discrete}), $e_N$ oscillates with $N$ for the DREM. What Figure \ref{Fig3} shows is that it also oscillates in the case of the tree.

The fact that the sizes of  the blocks in the replica approach may fluctuate  is a rather general phenomenon. It  could even occur  for continuous distributions of energies.  In fact  the expression (\ref{eq3},\ref{Yk-cont}) is a direct consequence of the fact that the  energies  close to the ground state energy have a distribution  well approximated by  a single   exponential distribution (\ref{PoissonP}). If for some  reason, for example near a phase transition,  this distribution would instead  be  well approximated by the   sum of two exponentials
$$\rho(E) \simeq A \,  e^{\beta_c (E-N \epsilon_c)} + A'  \, e^{\beta_c' (E-N \epsilon_c)} $$
then the function  $\phi(t)$ in (\ref{phi-continuous}) would become
$$\phi(t)=- \frac{A  \ e^{-N \beta_c \epsilon_c}} {\beta_c} \ t^\frac{\beta_c }{ \beta} \ \Gamma\left(1-\frac{\beta_c }{ \beta} \right)
- \frac{A'  \ e^{-N \beta_c '\epsilon_c}} {\beta_c'} \ t^\frac{\beta_c' }{ \beta} \ \Gamma\left(1-\frac{\beta_c' }{ \beta} \right)
$$
Then one could repeat the calculation of section (\ref{disc1}) and obtain that
\begin{equation}
\langle Y_k (\beta) \rangle = 
W \frac{\Gamma(k-\mu) }{ \Gamma(k) \, \Gamma(1-\mu)} + 
W' \frac{\Gamma(k-\mu') }{ \Gamma(k) \, \Gamma(1-\mu')}  
\label{Yk2}
\end{equation}
with $\mu=\frac{\beta_c }{ \beta}$ ,  $\mu'=\frac{\beta_c' }{ \beta}$ and  the weights
given by

\begin{align*}
	 W= & \frac{A  \ e^{-N \beta_c \epsilon_c}} {\beta} \
	\Gamma\left(1-\frac{\beta_c }{ \beta} \right)
	\int_0^\infty t^{\frac{\beta_c }{ \beta}-1} \ e^{\phi(t)} dt 
	\ \ \ ; \\ \ \ \ 
	W'= &
	\frac{A'  \ e^{-N \beta_c' \epsilon_c}} {\beta} \
	\Gamma\left(1-\frac{\beta_c' }{ \beta} \right)
	\int_0^\infty t^{\frac{\beta_c' }{ \beta}-1} \ e^{\phi(t)} dt .
\end{align*}
Clearly (\ref{Yk2}) shows that the sizes of the blocks in the replica calculation take the value $\mu$ with probability $W$ and $\mu'$ with probability $W'$ (it is easy to check that $W$ and $W'$ are positive numbers and that
$W+W'=1$ using the fact that $\phi(0)=0$ and $\phi(\infty))=-\infty$)).

As  the discrete energy case can be viewed as  a situation where the  density of energies is a sum of exponentials with complex growth rates, it is not surprising that we found boxes with complex sizes in (\ref{Yk-complex}).

\bibliographystyle{unsrtnat}
\bibliography{pjm-master}

 \end{document}